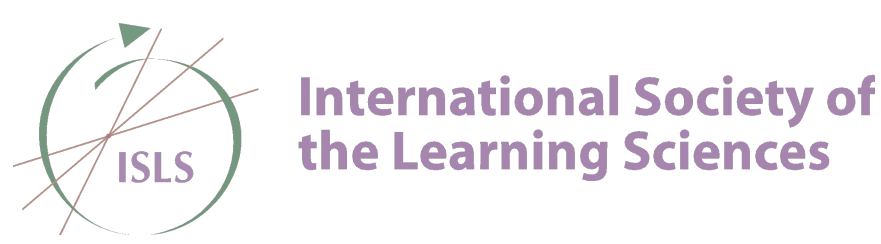

# MIRACLE: Multi-Agent Intelligent Regulation to Advance Collaborative Learning Environment

Shuang Li, Haiyang Xin, Yimeng Sun, Qiannan Niu
lishuang@cocorobo.cc, Tony@cocorobo.cc, sunyimeng@cocorobo.cc, niuqiannan@cocorobo.cc,
COCOROBO Limited
Lingyun Huang, The Education University of Hong Kong, lingyunhuang@eduhk.hk
Gaowei Chen, The University of Hong Kong, gwchen@hku.hk
Yibing Zhang, Nanjing Normal University, zyc@njnu.edu.cn
Ching Sing Chai, The Chinese University of Hong Kong, CSChai@cuhk.edu.hk

**Abstract:** Effective collaboration requires Socially Shared Regulation (SSRL), but students often lack these skills. This study introduces the MIRACLE (Multi-Agent Intelligent Regulation to Advance Collaborative Learning Environment) system, which supports SSRL by orchestrating metacognitive regulation and proactively providing emotional and motivational support. We conducted a quasi-experimental study with 90 fifth-grade students. The experimental group (n=42) used a collaborative platform CocoNote equipped with MIRACLE, while the control group (n=48) used the same platform with a general GPT assistant. Quantitative results show the MIRACLE group achieved significant gains across SSRL phases (Planning, Monitoring, Reflection) and produced higher-quality collaborative artifacts compared to the control group. Qualitative findings indicate students perceived MIRACLE as an effective facilitator for cognitive, regulatory, and emotional support. This study demonstrates that specialized, orchestrated AI systems are more effective than generic AI in enhancing SSRL.

## Introduction

Collaboration has been widely recognized as a powerful pedagogical approach that enhances students' academic performance, knowledge construction, and problem-solving skills (Miller & Hadwin, 2024). However, effective collaboration is far more complex than individual learning and is often hindered by challenges such as unequal participation, dominance of certain members, and difficulties in reaching consensus (Zheng et al., 2023). A major underlying reason for these challenges is the insufficient development of Socially Shared Regulation of Learning (SSRL)—the collective process through which group members plan, monitor, and regulate their joint learning activities (Järvenoja et al., 2020). While SSRL is essential for productive collaboration, many learners lack the regulatory competence to coordinate their efforts effectively, thereby requiring structured guidance and external support (Järvelä et al., 2013).

Earlier research on technology-enhanced collaborative learning provided such structured support through collaboration scripts and group awareness tools (Järvelä et al., 2015; Miller & Hadwin, 2015). With the rise of Generative Artificial Intelligence (GenAI), new opportunities have emerged to offer more flexible and context-aware support. Leveraging GenAI's strengths in language understanding and dynamic feedback, researchers have begun developing conversational agents capable of delivering personalized and "in-the-moment" guidance (Kasneci et al., 2023). Edwards (2024) introduced a proactive speech agent that prompts students to reflect and regulate during group discussions. While the proactive agent demonstrates potential, it provides limited support types and may not respond adequately to students' diverse and evolving needs. Recent multi-agent systems (Xin et al., 2025) attempt to expand functionality through the integration of multiple agents (e.g., proactive and reactive agents). However, this multi-agent configuration introduces a new challenge: uncoordinated interactions with multiple agents can increase students' cognitive load and lead to fragmented or inconsistent learning experiences. Thus, there remains a pressing need for intelligently orchestrated agent systems that can dynamically coordinate different regulatory functions in a structured and seamless manner.

To address this gap, this study presents MIRACLE (Multi-Agent Intelligent Regulation to Advance Collaborative Learning Environments)—an orchestrated multi-agent system designed to enhance SSRL by combining metacognitive regulation with proactive emotional and motivational support. Implemented within a collaborative learning platform CocoNote, MIRACLE aims to provide students with adaptive and structured assistance. This study investigates:

RQ1: Does the MIRACLE system enhance students' SSRL abilities and collaborative learning outcomes?

RQ2: How do students perceive the MIRACLE system and its influence on their collaborative learning?

## Theoretical Foundations

### Socially Shared Regulation of Learning

Derived from self-regulated learning (SRL) theory, SSRL is the process through which group members consciously regulate their collective cognitive, behavioral, motivational, and emotional states. This management occurs through an iterative cycle of planning, monitoring, evaluating, and regulating phases (Hadwin et al., 2017). The MIRACLE system is engineered to foster and maintain this SSRL process by delivering regulatory support for the planning, monitoring, and reflection stages, alongside cognitive assistance during task execution.

### Trigger Regulation Framework

The Trigger Regulation Framework (Järvelä & Hadwin, 2024) identifies triggering events as critical moments for supporting learners' collaboration. These triggering events refer to circumstances or behavioral patterns that impede learning progress, such as diminished participation or adverse emotional states, which present valuable opportunities for facilitating metacognitive development (Edwards et al., 2024). Drawing upon this framework, the MIRACLE system's proactive agents were designed to promote collective metacognitive awareness and participation through the provision of metacognitive, socio-emotional, and motivational scaffolding.

## The Collaborative Learning Platform CocoNote

The collaborative learning platform CocoNote integrates a hypermedia whiteboard, a group chatroom, and the MIRACLE system to facilitate SSRL. The whiteboard allows learners to co-construct knowledge by creating and connecting multimodal notes (text, images, and videos), while the chatroom enables synchronous communication for task deliberation and coordination.

### The MIRACLE Agent System Design

The MIRACLE system (implemented within the platform) is powered by GPT-5-Nano and incorporates both reactive agents and one proactive agent to support the SSRL process.

*Reactive Agents*. The reactive agents employ a hierarchical agent architecture built using the LangGraph framework (LangChain, 2024). At its core is a Boss Agent that orchestrates all student-agent interactions by intelligently routing student requests to four specialized agents: the Planning Agent, Monitoring Agent, Reflection Agent, and Knowledge Agent (see Figure 1). When students mention the Boss Agent in the chatroom, it analyzes their messages to determine whether they need knowledge support or metacognitive guidance, then directs them to the appropriate specialized agent. All agents access shared real-time data from both the chatroom and whiteboard, enabling contextually grounded responses. This design allows students to focus on collaborative tasks rather than navigating between different agents.

**Figure 1**
*The Intelligent Agent Orchestration Architecture*

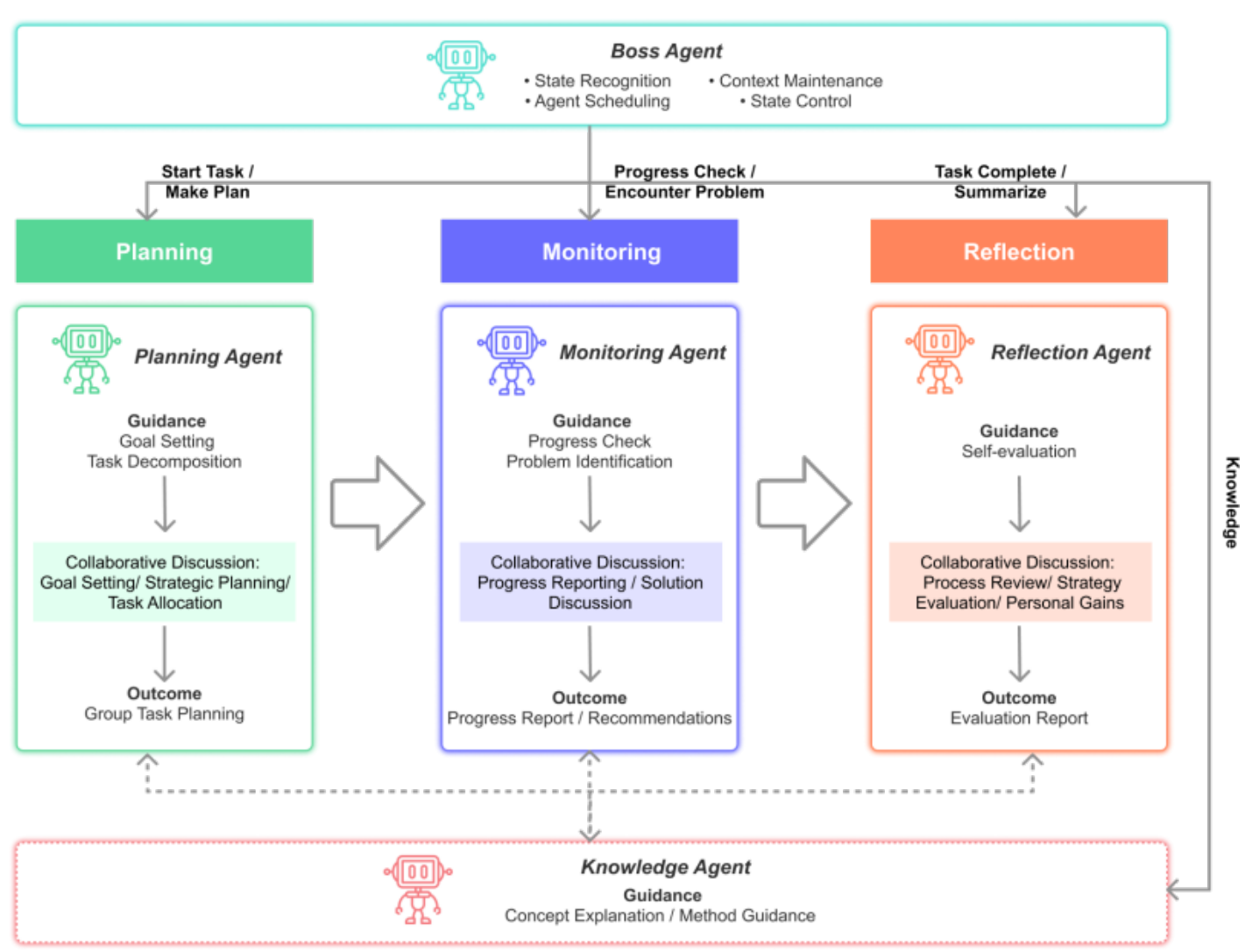

*Proactive Agent.* Recognizing the critical role of emotional and motivational factors in sustaining collaborative engagement, particularly among younger learners, the MIRACLE system also features a dedicated Lightbulb Agent that operates independently from the metacognitive support agents. The Lightbulb monitors both chatroom conversations and whiteboard activities to detect triggering events that signal emotional or motivational challenges, such as extended periods of individual inactivity, expressions of confusion or frustration, declining participation rates, or absence of progress in task execution. Unlike reactive agents that respond to explicit student requests, the Lightbulb operates proactively, initiating interventions when triggering events are detected. Upon identifying a trigger, the agent provides feedback via an interface and notifies learners by flashing a "lightbulb" icon on the top-right corner of the whiteboard. Once learners view the prompt, the lightbulb stops flashing until the next trigger.

## Research Design and Methods

This quasi-experimental study involved 90 fifth-grade students (aged 10–11) from two parallel classes in an elementary school from a coastal city with well-developed educational infrastructure in Southern China. The experimental group (n = 42) used the collaborative platform equipped with the MIRACLE system, while the control group (n = 48) used the same platform—with identical whiteboard and chatroom features—but interacted with a general GPT assistant (also powered by GPT-5-Nano) instead of MIRACLE. The GPT assistant received a general-purpose system prompt without specialized pedagogical instructions and operated in a purely reactive mode, responding only when students mentioned it in the chatroom; it had no proactive triggering mechanism or orchestrated agent routing. Students worked in groups of 4–6 to complete a 120-minute design-thinking project in the computer lab, which required them to plan and present an innovative gift design, including its concept, appearance, materials, and production process. Multiple data sources were collected to ensure validity. The Socially Shared Regulated Learning Questionnaire (Yang et al., 2025) was administered before and after the intervention to assess students' SSRL abilities. Group artifacts were collected and evaluated for quality, and six groups (three from each condition) participated in post-study interviews to provide qualitative insights into their collaborative experiences.

### Data Treatment

To examine intervention effects on SSRL abilities (RQ1), we performed 2×2 repeated-measures ANOVAs on four SSRL dimensions (Yang et al., 2025), with paired t-tests for within-group changes. Two raters assessed artifacts (N=18) using a three-dimensional rubric (ICC=.89). Semi-structured interviews with six groups (three experimental, three control) were analyzed using reflexive thematic analysis (Cohen's κ =.80).

## Findings

### RQ1. Does the MIRACLE system enhance students' SSRL abilities and collaborative learning outcomes?

**Table 1**

*Pre-Post Changes and Intervention Effects on SSRL Dimensions*

| Dimension | Group | Pre-test | Post-test | Within-Group Change | | Time × Group Interaction | | |
|---|---|---|---|---|---|---|---|---|
| | | M ± SD | M ± SD | t | p | F | p | $\eta^2p$ |
| **Planning** | Experimental | 26.43±5.92 | 28.38±4.38 | -2.07 | .045* | 7.44 | .008** | .078 |
| | Control | 25.75±5.41 | 24.19±6.26 | 1.78 | .082 | | | |
| **Monitoring** | Experimental | 24.48±6.81 | 28.43±4.79 | -3.09 | .004** | 7.15 | .009** | .075 |
| | Control | 25.46±6.30 | 25.15±7.55 | 0.32 | .753 | | | |
| **Reflection** | Experimental | 24.79±5.56 | 27.10±4.80 | -2.28 | .028* | 3.21 | .077† | .035 |
| | Control | 23.96±5.15 | 24.06±5.96 | -0.14 | .888 | | | |
| **Emotion** | Experimental | 46.14±8.62 | 47.45±6.54 | -0.92 | .365 | 1.45 | .231 | .016 |
| | Control | 45.08±7.03 | 44.27±8.87 | 0.76 | .452 | | | |

Repeated measures ANOVA revealed significant Time × Group interactions for Planning and Monitoring (see Table 1). The experimental group improved significantly on Planning, Monitoring, and Reflection, while the control group showed no changes. Emotion Regulation showed no significant effects.

Mann-Whitney U tests were conducted to compare artifact quality between experimental and control groups, as the data violated normality assumptions (Shapiro-Wilk test, $p < .05$). Results revealed that experimental groups (Mdn = 7.50, M = 7.61, SD = 0.99) produced significantly higher-quality artifacts than control groups (Mdn = 5.50, M = 4.78, SD = 1.48), $z = -3.27$, $p < .001$, $r = .77$, representing a large effect size. Dimensional analyses showed that experimental groups outperformed control groups across all three aspects.

## RQ2.How do students perceive the MIRACLE system and its influence on their collaborative learning?

Semi-structured interviews with six groups revealed how students perceived the multi-agent system's influence on their collaborative learning experiences.

### Boss Agent Orchestration: Cognitive Support and Process Regulation

Experimental students perceived the Boss Agent as a holistic facilitator that streamlined both their learning process and their group collaboration. Its value spanned the entire SSRL cycle, manifesting in distinct roles at each stage. For initiating the task, it served as both an ideation partner for planning and an on-demand expert for knowledge support. This support then transitioned into a supervisory role during task execution. For monitoring, the agent acted as an objective third-party supervisor, prompting adjustments when it "said our progress wasn't particularly good... we took the suggestions and things got much better" (Student C). For reflection, it functioned as an administrator, providing comprehensive summaries that students "felt... was very thorough" (Student D).

Crucially, students reported that this automated, objective support was the key mechanism for unifying group decisions and reducing conflicts. By acting as a central coordinator for the team's SSRL process, the Boss Agent made the collaboration more orderly, experienced by students as seamless assistance that "really sped up our task progress" (Student C).

### Lightbulb Agent: Emotional Support and Member Activation

Students generally perceived the Lightbulb Agent as providing valuable emotional support rather than intrusion. Students appreciated its personalized nature, noting that its suggestions were highly relevant to their specific contributions: "The Lightbulb knows what I have made on the white board, its suggestion is relevant to my actual situation" (Student E).

Most notably, the Lightbulb's visible interventions provided crucial socio-emotional scaffolding, fostering a sense of unity and activating disengaged members. Its prompts were seen as warm encouragement that helped students engage in teamwork: "It told me no need to be anxious and just take my time" (Student F), and motivated reluctant participants to rejoin the collaborative effort: "At first, some people in our group weren't doing anything. But when they saw the Lightbulb... they started cooperating with us" (Student G).

### Contrasting Experiences and Overall Impact

The positive perceptions of the MIRACLE system stand in stark contrast to the experiences of the control group. Experimental students unanimously affirmed that their collaboration improved: "Yes, it got better... things became more organized" (Student H).

In contrast, control group students reported significant coordination failures: "It didn't go well... mainly because we weren't united, and the group leader we chose didn't have management ability" (Student I). Notably, their primary desire was for basic oversight mechanisms—"The platform should remind or report students who go off-topic" (Student J). This divergence suggests that MIRACLE satisfied foundational regulatory needs, allowing groups to focus on higher-level collaboration.

## Discussion, Limitations and Future Directions

The experimental group's consistent gains in Planning, Monitoring, and Reflection—the three core phases of SSRL—demonstrate that MIRACLE's pedagogically intelligent and orchestrated agents effectively scaffolded the entire regulatory process. In contrast, the control group, despite using a general GPT assistant within the same collaborative environment, showed little progress, reinforcing the necessity of structured, pedagogical integration over generic AI assistance for fostering meaningful regulation in collaboration. The lack of significant quantitative gains in Emotion Regulation, despite qualitative evidence of emotional support, likely reflects that emotional regulation competencies require longer internalization periods than the 120-minute intervention provided. The study's small, age-specific sample and short intervention duration limit generalizability and long-term inference. Future research should employ larger samples, longer interventions, and multimodal data to deepen understanding of pedagogical AI's impact on collaborative regulation.